\documentstyle[12pt]{article}
\def\be{\begin{eqnarray}}
\def\ee{\end{eqnarray}}
\begin{document}
 \thispagestyle{empty}
 \begin{flushright}
 {MZ-TH/94-02}    \\
 {January 1994}           \\
 \end{flushright}
 \vspace{1cm}
 \begin{center}
 {\bf \large
A Precise Determination of $\tan\beta$ from Heavy Charged Higgs Decay}
 \end{center}
 \vspace{1cm}
 \begin{center}
 {Andrzej Czarnecki\footnote{Address after July 1st, 1994: Institut f\"ur
Theoretische Teilchenphysik, Universit\"at
Karlsruhe, D-76128 Karlsruhe, Germany}}\\
  \vspace{.3cm}
 {\em Institut f\"ur Physik, Johannes Gutenberg-Universit\"at,\\
 \mbox{D-55099} Mainz, Germany}
 \end{center}
 \begin{center}
 and
 \end{center}
 \begin{center}
 {James L.~Pinfold}\\
 \vspace{.3cm}
 {\em Department of Physics, University of Alberta, \\
      Edmonton, Canada T6G 2J1}
 \end{center}
 \hspace{3in}
 \begin{abstract}
 We compute the energy spectrum of charged leptons in the decay
$H^+\to \bar b+(t\to bl\nu_l)$.  The shape of the lepton spectrum
obtained, and also the mean lepton energy, are sensitive to the
handedness of the intermediate top quark. This sensitivity can be used
to precisely determine $\tan\beta$, a fundamental parameter of two
Higgs doublet models.
 \end{abstract}

 \vspace{15mm}

 \begin{center} PACS numbers:  14.80.Gt, 12.15.Cc
 \end{center}

 \newpage

Precision measurements of electroweak parameters and the strong
coupling constant ($\alpha_s$) at LEP \cite{jim1} are completely
consistent with the minimal supersymmetric extension to the standard
model (MSSM), with a MSSM mass scale of around 1 TeV \cite{jim2}.

In the MSSM the single standard model (SM) complex scalar doublet, or
Higgs doublet, is replaced by two Higgs doublets (for a review
see ref.~\cite{hunter}). After spontaneous
symmetry breaking the MSSM Higgs sector consists of: two CP-even
bosons $h^0$ and $H^0$ ($m_{h^0} < m_{H^0}$); a CP-odd boson $A^0$;
and two charged scalars $H^\pm$. The presence of a Higgs sector that
includes two doublets is a feature of most supersymmetric models. In
models with only Higgs doublets (and singlets) there is no coupling of
the charged Higgs to vector boson pairs ($WZ$ or $W\gamma$) at tree
level.
The most important decay channels of the $H^\pm$ are into fermion pairs
or into $W^\pm h^0$. However, $H^\pm$ decays into states involving charginos
can, in some regions of the parameter space, be quite large
\cite{Barnett}. For a $H^\pm$
with mass ($m_{H^\pm}$) greater than the sum of top and bottom quarks
($m_t+m_b$) the
dominant decay process is $H^+ \to t\bar b$. A future linear $e^+e^-$
collider would be an ideal place to search for the decay
$H^+ \to t\bar b$. However, in practice, planned machines are limited
to beam energies around 500 GeV. Recent studies
\cite{gunion93,barger93} indicate that it is feasible to also search
for a $H^\pm$ (with $m_{H^\pm}> m_t$) at hadron colliders, such as the
LHC, over a large region of the available parameter space.

In the present letter we observe that a
 charged Higgs boson of the Minimal Supersymmetric Model
significantly heavier than the top
quark offers an opportunity to precisely measure the fundamental
parameter of a two Higgs doublet model, namely the ratio of vacuum
expectation values of the two doublets, $\tan\beta\equiv v_2/v_1$.
The reason for this lies in the structure of the coupling of the
charged Higgs boson to $tb$ quarks:
\be
{\cal L}={g V_{tb} \over \sqrt{2}m_W} H^+\bar t\left[ m_t \cot\beta L
+ m_b \tan\beta R\right]b + H.c.,
\ee
where $R,L=(1\pm \gamma_5)/2$, $m_W$ is the mass of the $W$ boson, $g$
is the weak coupling constant, and $V_{tb}$ is the relevant element of
the CKM matrix. From this Lagrangian we can see that
the ratio of numbers of
left and right polarized top quarks produced in the decay
$H^+ \to t\bar b$ depends on the coefficients of left and right chiral
projection operators $L, R$. At the tree level:
\be
{N_L\over N_R}=\left( {m_b\over m_t} \right)^2 \tan^4\beta,
\label{nlnr}
\ee
where $N_{L(R)}$ are the partial rates of production of
left (right) handed top quarks.
If $H^+$ is significantly heavier than the top quark, so that in
its decay the top is emitted with large velocity, we can determine the
handedness  of the top quark by analyzing the energy spectrum of the
charged lepton produced in the semileptonic decay of $t$ \cite{peskin}.
We see from the equation (\ref{nlnr}) that the ratio $N_L/N_R$ is very
sensitive to the value of $\tan\beta$. However, in practice this means
that using this method one will only be able to determine the precise
value of $\tan\beta$ over a limited range.
For extreme
values of $\tan\beta$ the top quark will be produced exlusively in one
polarization state, $t_R$ for small $\tan\beta$ and  $t_L$ for large.
Examination of the lepton spectrum will then be useful to determine
the order of magnitude of $\tan\beta$.

It is well known \cite{mend90,mend91,lioakes91,cd93}
that the decay width of the charged Higgs
boson is strongly modified by QCD corrections. These
corrections will also influence  equation (\ref{nlnr}). However, the
dominant part of corrections can be absorbed by replacing $m_b$ by the
running mass of $b$ ($\overline{m}_b$)
at the energy scale of the mass of $H^+$.
We use the threshold condition
$m_b(2m_b)=4.7$ GeV, which gives $\overline{m}_b\approx 3.6$ GeV at the
typical mass scale of the $H^+$ of a few hundred GeV \cite{gunion93}.
The corresponding effect on the mass of the top quark is of the order
of $5\%$. In our numerical estimates we take
$\overline{m}_t=150$ GeV and
$m_{H^+}\equiv m_H=300$ GeV.

In order to explicitly evaluate the energy distribution of leptons in
the decay chain $H^+\to \bar b+(t\to b+(W^+\to \bar l\nu_l))$ we
adopt the narrow
width approximation for both intermediate particles $t$ and $W^+$. The
total width of this decay is then
\be
\Gamma(H^+\to \bar bb\bar l\nu_l)
&=& {\Gamma(H^+\to t\bar b)\Gamma(t\to bW^+)\Gamma(W^+\to \bar l\nu_l)
\over \Gamma_t \Gamma_W}
\nonumber \\
&=&
{G_F^3\over 128\sqrt{2} \pi^3}
{m_W^3\over \Gamma_t \Gamma_W \overline{m}_t^3 m_H^3}
\left(\overline{m}_t^2\cot^2\beta +\overline{m}_b^2\tan^2\beta \right)
\nonumber \\
&&\times\left(m_H^2-\overline{m}_t^2 \right)^2
\left(\overline{m}_t^2-m_W^2 \right)^2
\left(2m_W^2+\overline{m}_t^2 \right),
\ee
where we have neglected mass of $b$ everywhere except in the coupling.
It can be seen that the total width of $H^+$  depends on
$\tan\beta$. However it does not distinguish between  the two values:
$\tan^2\beta=X$ and $\tan^2\beta=\overline{m}_t^2/
(\overline{m}_b^2X)$. Much more
information can be gained from the partial decay width $d\Gamma/dx$,
with $x\equiv 2E_l/m_H$ denoting the scaled energy of the lepton
in the rest frame of the $H^+$.
We introduce the following dimensionless variables
\be
u={\overline{m}_t^2\over m_H^2}, &\qquad & y={m_W^2\over m_H^2},
\nonumber\\
a={\overline{m}_b^2\over m_H^2}\tan^2\beta, &\qquad &
b={\overline{m}_t^2\over m_H^2}\cot^2\beta.
\ee
The kinematic limits of lepton energy are expressed by
\be
y\leq x \leq 1.
\ee

In order to compute the energy distribution we use
\be
d\Gamma={1\over 2m_H}(2\pi)^{-8}|{\cal M}|^2
dR_4\left(H;b,q,l,\nu\right),
\ee
where ${\cal M}$ is the matrix element and $R_4\left(H;b,q,l,\nu\right)$
is the 4 body
phase space with letters denoting 4-momenta of the corresponding
particles, and since we have to distinguish between the two $b$
quarks, we assign the letter $b$ to the one produced in the primary
decay $H^+\to t\bar b$ and the letter $q$ to the product of the top
quark decay. We decompose the 4 body phase space
in a manner similar to the calculation of QCD corrections
to the lepton energy spectrum in semileptonic top quark decays
\cite{jk1}
\be
dR_4\left(H;b,q,l,\nu\right)=dR_3\left(H;P,l,\nu\right)dz
dR_2(P;b,q),
\ee
where $P\equiv b+q$ is the 4-momentum of the two $b$ quarks, and $z$
is the square of their invariant mass, $z\equiv P^2/m_H^2$.
In the narrow width approximation the phase space integration is
simplified by replacing the propagators of $t$  and $W$ by delta
functions, and we end up with the following formula for the energy
distribution of leptons
\be
{d\Gamma\over dx}={3G_F^3\over 64 \sqrt{2} \pi^3}
{m_W^3 m_H^7\over \overline{m}_t \Gamma_t \Gamma_W} f(x)
\ee
where the dimensionless function $f(x)$ defined by
\be
f(x)=\left\{ \begin{array}{ll}
2ux(a-b)\left[x-y-u\ln(x/y)\right] +(a-bu)(x-y)(2u-x-y) &
\\
&  \hspace{-3cm} \mbox{if $  y\leq
x\leq \min(u,y/u) $}\\
2u^2x(a-b)(1-x+\ln x)+u^2(a-bu)(1-x)^2 &
\\
&  \hspace{-3cm} \mbox{if $  \max(u,y/u)\leq
x\leq              1  $}
\end{array}
\right.
\label{fdef1}
\ee
For values of $x$ between $u$ and $y/u$ we have to distinguish two
cases:
\be
f(x)=\left\{ \begin{array}{ll}
2ux(a-b)\left[u-y+u\ln(y/u)\right]+(a-bu)(y-u)^2      &
\\
&  \hspace{-2cm} \mbox{if $  u<y/u$}
\\
2ux(a-b)\left[x(1-u)+u\ln u\right]+x(a-bu)(1-u)(2u-x-xu)
&
\\
&  \hspace{-2cm} \mbox{if $ u>y/u $}
\end{array}
\right.
\label{fdef2}
\ee
We use these formulae to compute the mean scaled energy of leptons,
$\bar x$. The result is the same in both mass cases:
\be
\bar x={a(1+2u)+b(2+u)\over a+b}{u^2+2uy+3y^2\over 6u(u+2y)}.
\label{xmean}
\ee
We see from this formula that
if mass of the top quark approaches mass of the decaying Higgs boson
($u\rightarrow 1$) the mean energy looses sensitivity to values of $a$
and $b$. Therefore the determination of $\tan\beta$ using this method
will be better if the charged Higgs boson is significantly heavier
than the top quark, i.e. when the top produced in its decay has high velocity.

In order to illustrate the sensitivity of the lepton energy
distribution to the value of $\tan\beta$ we show in fig.~1 plots of
$f(x)$  for
two extreme cases, $\tan\beta=0.5$ (solid line)
and for $\tan\beta=83.33$ (dashed) (these two values correspond to the
same total decay rate).
We obtain the characteristic spectra of leptons from totally polarized
right and left handed top quarks \cite{peskin}.
In figure 2 we plot the dependence of the mean energy of leptons on
the value of $\tan\beta$. Provided the charged Higgs is significantly
heavier than the top quark, we obtain good sensitivity to
the precise value of $\tan\beta$ in the range $2\le \tan\beta \le 10$.
Outside of this range, the spectrum of leptons can
only serve to determine
whether $\tan\beta$ is small or large.

\section*{Acknowledgment}
We are grateful to Marek Je\.zabek and Apostolos Pilaftsis
for helpful discussions, and to Stefan Balk and Lars Bruecher for
sharing their computer expertise. This work
was supported by Deutsche Forschungsgemeinschaft (Germany).

\section*{Figure captions}
\begin{itemize}
\item[]Figure 1: Distribution of lepton energy in the decay $H^+\to
\bar b b\bar l\nu$, according to equations
(\ref{fdef1}-\ref{fdef2}), calculated for $m_H$ =
300 GeV, $m_W$ = 80.22 GeV, for two values of $\tan\beta$.
\item[]Figure 2: Sensitivity of the mean scaled energy of leptons
(see eq.~\ref{xmean}) to the value of $\tan\beta$, shown for
three values of mass of $H^+$.
\end{itemize}

\begin{thebibliography}{99}

\bibitem{jim1}
For example, T. Kawamoto, Aspen Winter Conference in Particle Physics,
Aspen, Colorado, January 1993; R. Tanaka, Int. Conf. on High Energy
Physics, Dallas, Texas, Aug. 1992;
T. Wyatt,  Int. Conf. on High Energy
Physics, Dallas, Texas, Aug. 1992.

\bibitem{jim2}
P. Langacker, {\em Electroweak Physics Beyond the Standard Model},
UPR-0492T. Invited talk given at the Int. Workshop on
Electroweak Physics Beyond the Standard Model, Valencia, Spain, Oct. 1991.


\bibitem{hunter}
J.F. Gunion, H.E. Haber, G.~Kane, and S.~Dawson.
\newblock {\em The {Higgs} Hunter's Guide}.
\newblock Addison-Wesley, Redwood City, {CA}, 1990.
\newblock Errata: hep-ph/9302272.

\bibitem{Barnett} R.M. Barnett, J.F. Gunion and H.E. Haber, in
Proc.~of the 1988 University of California at Davis Workshop
on Intermediate Mass and Non-Minimal Higgs Bosons, J.F.~Gunion and
L.~Roszkowski (eds.).


\bibitem{gunion93}
J.F. Gunion,
\newblock {\em Detecting
 the {$tb$} decays of a charged {H}iggs boson at a hadron
  supercollider},
\newblock preprint UCD-93-40.

\bibitem{barger93}
V.~Barger, R.J.N. Phillips, and D.P. Roy,
\newblock {\em Heavy charged {H}iggs signals at the {LHC}},
\newblock hep-ph/9311372.

\bibitem{peskin}
C.R. Schmidt and M.E. Peskin,
\newblock  Phys. Rev. Lett. {\bf 69} (1992) 410.

\bibitem{mend90}
A.~M\'endez and A.~Pomarol,
\newblock Phys. Lett. {\bf B252} (1990) 461.

\bibitem{mend91}
A.~M\'endez and A.~Pomarol,
\newblock  Phys. Lett. {\bf B265} (1991) 177.

\bibitem{lioakes91}
C.S. Li and R.J. Oakes,
\newblock  Phys. Rev.  {\bf D43} (1991) 855.

\bibitem{cd93}
A.~Czarnecki and A.I. Davydychev,
\newblock {\em Two-loop renormalization group analysis of hadronic decays of a
  charged {H}iggs boson},
\newblock Alberta Thy-38-93.

\bibitem{jk1}
M.~Je{\.z}abek and J.~H. K{\"u}hn,
\newblock  Nucl. Phys. {\bf B320} (1989) 20.

\end{thebibliography}
\end{document}